\documentclass[twocolumn,showpacs,preprintnumbers,
amsmath,amssymb,aps,prd,nofootinbib,superscriptaddress,
eqsecnum]{revtex4}
\usepackage{graphicx,color,flafter}

\makeatletter
\renewcommand{\p@subsection}{}
\makeatother

\newcommand{\del}{\partial}

\def\sl#1{\setbox0=\hbox{$#1$}               
   \dimen0=\wd0                                 
   \setbox1=\hbox{/} \dimen1=\wd1               
   \ifdim\dimen0>\dimen1                        
      \rlap{\hbox to \dimen0{\hfil/\hfil}}      
      #1                                        
   \else                                        
      \rlap{\hbox to \dimen1{\hfil$#1$\hfil}}   
      /                                         
   \fi}

\begin{document}

\title{Kurtosis and compressibility near the chiral phase transition}
\author{B.~Stoki\'c}
\affiliation{%
Gesellschaft f\"ur Schwerionenforschung, GSI,  D-64291 Darmstadt, Germany}
\author{B.~Friman}
\affiliation{%
Gesellschaft f\"ur Schwerionenforschung, GSI,  D-64291 Darmstadt, Germany}
\author{K.~Redlich}
\affiliation{%
Institute of Theoretical Physics, University of Wroclaw, PL--50204 Wroc\l aw, Poland}
\affiliation{%
Technical University Darmstadt, D-64289 Darmstadt, Germany}

\pacs{}

\date{\today}

\begin{abstract}

The properties of net quark number fluctuations in the vicinity of the QCD chiral phase
transition are discussed in terms of an effective chiral model in the mean-field approximation. We focus on the ratio of the  fourth- to second-
order cumulants (kurtosis) and the compressibility of the system and discuss
their dependence on the pion mass. It is shown that near the chiral phase transition, both observables
are sensitive to the value of $m_\pi$. For physical $m_\pi$, the kurtosis exhibits a peak
whereas the inverse compressibility shows a dip at the pseudocritical temperature. These structures
disappear for large  $m_\pi$. Our results, obtained in an effective model with two flavors, are qualitatively consistent with recent results of 2+1 flavor lattice gauge theory. We also discuss the high- and low-temperature properties of these observables and  the role of  the coupling of the quark degrees of freedom to the Polyakov loop.

\end{abstract}

\pacs{}

\setcounter{footnote}{0}

\maketitle

\section{Introduction}

Understanding the critical properties of the QCD medium and its phase diagram is a central problem in strong interaction physics, which is being addressed in
both theoretical and experimental studies. Lattice Gauge Theory (LGT), in particular, has provided
substantial insight into the physics of the transition between the low temperature hadronic
and high temperature quark--gluon plasma phase
\cite{lattice:ejiri,Allton:2003vx,fodor,lgtm,lgtp,gupta,FN,Ejiri:2005wq,F1}. Effective models also provide a useful qualitative  description of the chiral phase transition
\cite{Ratti:2006wg,Fukushima,sfr:prd,Ghosh:2006qh,Schaefer:PQM}. Here, the expectation that the critical behavior of such models is governed by the same universality class as QCD \cite{pisarski} is of central importance. Furthermore, recent experimental data from
heavy-ion collisions at RHIC and SPS provide new insight into the properties and behavior
of a deconfined medium created in dynamical systems \cite{proc}.

The phase diagram of a thermodynamic system may be mapped out by studying  fluctuations and their
response to changes of the thermodynamic parameters. In QCD the fluctuations of conserved
charges are particularly pertinent for exploring deconfinement and chiral symmetry restoration
\cite{Allton:2003vx,Ejiri:2005wq,hatta,stephanov1,stephanov2,fluct}.

Studies of QCD on the lattice at finite temperature and density show that  the ratio
of the quartic to quadratic fluctuations of the net quark number $R_{4,2}$ (kurtosis~\footnote{Strictly speaking, in statistics kurtosis is given by the fourth moment divided by the second moment squared, i.e., $\langle (\delta N)^4\rangle/\langle (\delta N)^2\rangle^2$. In this letter we employ the so called kurtosis excess, with an atypical normalization $R_{4,2}=\langle (\delta N)^4\rangle/\langle (\delta N)^2\rangle - 3\langle (\delta N)^2\rangle$. It turns out that this ratio is useful for distinguishing the confined and deconfined phases~\cite{Ejiri:2005wq}. In the interest of brevity, we sacrifice rigor and refer to $R_{4,2}$ as kurtosis.}) is a valuable
probe of deconfinement and chiral dynamics \cite{Ejiri:2005wq,FN,F1}. Furthermore, both
model calculations \cite{kunihiro,our} and LGT studies \cite{Allton:2003vx,FN,Ejiri:2005wq} indicate that the
inverse compressibility $R_\kappa$ is a useful observable for  identifying
the position of the critical end point CEP in the QCD phase diagram.

In the limits of high and low temperatures, $R_{4,2}$ depends on the net quark
content of the relevant baryon number carrying degrees of freedom~\cite{Ejiri:2005wq}. Consequently, one expects a strong reduction of $R_{4,2}$ at the confinement-deconfinement transition. Furthermore, the
inverse compressibility $R_\kappa$  vanishes at the CEP, due to the divergence of the net quark number
fluctuations at a critical point, belonging to the $Z(2)$ universality class \cite{hatta,stephanov1}. The first LGT
studies of these observables, which were performed in 2-flavor QCD with a pion mass  of  $m_\pi\simeq 770$ MeV
\cite{Allton:2003vx,Ejiri:2005wq}, confirmed the expected behavior of $R_{4,2}$ above and below the deconfinement transition. However, the results for the inverse compressibility $R_\kappa$ shows no sign of the expected decrease at the crossover
temperature $T_{pc}$; the LGT results for both $R_{4,2}$ and $R_\kappa$ interpolate smoothly between the limiting values in the hadronic and quark gluon plasma phases.

By contrast, results for $R_{4,2}$ and $R_\kappa$  
obtained recently in 2+1-flavor QCD with an almost physical pion mass, $m_\pi\simeq 220$ MeV, exhibit a qualitatively different temperature dependence~\cite{FN}. In these
calculations,  the kurtosis was found to develop a peak whereas the inverse compressibility acquired a dip
at the pseudocritical temperature. These results indicate that, close to the phase transition, these ratios depend on the strength of the explicit symmetry breaking term, i.e. on the value of the pion mass.  

In this letter, we use universality arguments, relating QCD with chiral effective models, to explore
the properties of the kurtosis and the inverse compressibility in hot and dense matter. We focus on
the dependence of $R_{4,2}$  and $R_\kappa$ on the pion mass near the chiral transition. As an
effective model of QCD we use the quark-meson Lagrangian coupled to the Polyakov loop in the mean field
approximation \cite{Schaefer:PQM}. 

We show, that the model results for $R_{4,2}$ and $R_\kappa$ are consistent with the LGT results and find that the peak/dip structures are indeed strongly dependent on the value of the pion mass. Given this, we conclude that the
non-monotonic behavior of these ratios found in LGT, for an almost physical pion mass, is due to the $O(4)$ chiral dynamics of 2-flavor QCD.  
Since the explicit symmetry breaking in the strange sector is large, one expects the strange quarks to decouple from the critical fluctuations and consequently that the critical properties, for physical quark masses, are dominated by the up-down sector. This observation supports the notion that the scaling properties of 2+1-flavor QCD are also governed by the $O(4)$ universality class.



\section{The chiral model}

In order to explore the influence of chiral dynamics on  the above observables we use the effective
Lagrangian of quark-meson model coupled to the Polyakov loop (PQM model)~\cite{Schaefer:PQM}
\begin{eqnarray}\label{eq:pqm_lagrangian}
  {\cal L} &=& \bar{q} \,(i\sl{D} - g (\sigma + i \gamma_5
  \vec \tau \vec \pi ))\,q
  +\frac 1 2 (\partial_\mu \sigma)^2+ \frac{ 1}{2}
  (\partial_\mu \vec \pi)^2
  \nonumber \\
  && \qquad - U(\sigma, \vec \pi )  -{\cal U}(\ell,\ell^{*})\ ,
\end{eqnarray}
with  ${\cal U}(\ell,\ell^{*})$ being an  effective potential of the gluon field expressed in terms
of the traced Polyakov loop $\ell$ and its conjugate $\ell^{*}$. The $O(4)$ representation of the meson fields is
$\phi=(\sigma,\vec{\pi})$ and the corresponding $SU(2)_L\otimes SU(2)_R$ chiral  representation is
given by $\sigma+i\vec{\tau}\cdot\vec{\pi}\gamma_5$. Consequently, there are $N_f^2=4$ mesonic degrees
of freedom coupled to $N_f=2$ flavors of constituent quarks $q$. The effect of gluons are accounted for by introducing an effective
field, which couples to the quarks by means of a covariant derivative
\begin{equation}
 D_{\mu}=\del_{\mu}-iA_{\mu}.
\end{equation}
Here the spatial components of the gauge field are chosen such that their average vanish,
i.e. $A_{\mu}=\delta_{\mu0}A_0$.

In the mean field approximation the thermodynamics of the model is described  by the thermodynamic potential
\cite{Schaefer:PQM}
\begin{eqnarray}  \Omega &=& {\mathcal U}(\ell,\ell^{*}) +
  U(\sigma) + \Omega_{\bar qq}(\ell,\ell^{*},\sigma),
\end{eqnarray}
with the gluonic, mesonic and quark/antiquark contributions, respectively.

The mesonic
part of the potential is of the form
\begin{equation}
U(\sigma,\vec{\pi})=\frac{\lambda}{4}\left(\sigma^2+\vec{\pi}^2-v^2\right)^2-h\sigma,
\end{equation}
where the last term describes the interaction of the scalar field $\sigma$ with an external field $h$. This term explicitly breaks the chiral symmetry of the Lagrangian and gives the pion a finite mass $m_\pi^2=h/\sigma$.

The gluonic contribution
to the thermodynamic potential is parametrized as an effective potential for the Polyakov loop and its conjugate
\begin{equation}
 \frac{{\cal U}(\ell,\ell^{*})}{T^4}=
-\frac{b_2(T)}{2}\ell^{*}\ell
-\frac{b_3}{6}(\ell^3 + \ell^{*3})
+\frac{b_4}{4}(\ell^{*}\ell)^2\,\label{eff_potential},
\end{equation}
where
\begin{eqnarray}
\hspace{-4ex}\label{coef}
  b_2(T) &=& a_0  + a_1 \left(\frac{T_0}{T}\right) + a_2
  \left(\frac{T_0}{T}\right)^2 + a_3 \left(\frac{T_0}{T}\right)^3.
\end{eqnarray}
The temperature independent coefficients, $a_i$ and $b_i$ in  Eqs. (\ref{eff_potential}) and
(\ref{coef})  have the following  values:
  $a_0 = 6.75$, $a_1 = -1.95$, $a_2 = 2.625$, $a_3 = -7.44$,
$b_3 = 0.75$ and $b_4 = 7.5$. They are chosen  so that the LGT results for the equation of state of a pure
$SU(3)$ gauge theory are reproduced~\cite{Ratti:2006wg}. Furthermore, for the parameter $T_0$ we choose $T_0=208$ MeV, the value favored in ref.~\cite{Schaefer:PQM} for the two-flavor sector.

The quark/antiquark contribution to the thermodynamic potential is given by
\begin{eqnarray} \label{eq:qqbar_pot}
\Omega_{\bar qq} =-P_{q\bar q}(T,\mu)= -2N_f T
\!\!\int\!\!\frac{d^3p}{(2\pi)^3}\hspace{3cm}&&  \\
\left\{\ln\!\! \left[1\! +\! 3 (\ell +
    \ell^* e^{-(E_p-\mu)/T})e^{-(E_p-\mu)/T}\!\!+\!\!e^{-3(E_p-\mu)/T}\right]
  +\right.&&\nonumber \\
 \left.  \ln\!\! \left[1\! +\! 3 ( \ell^* + \ell
    e^{-(E_p+\mu)/T})e^{-(E_p+\mu)/T}\!\!+\!\!e^{-3(E_p+\mu)/T}\right]
\right\},&&\nonumber
\end{eqnarray}
where $\mu$ is the quark chemical potential and $E_p=\sqrt{\vec{p}\,^2+m^2_q}$ is the  quasi-particle energy with $m_q=g\sigma$ being the
constituent quark mass.

In the mean field approximation the three independent fields $\sigma$, $l$ and $l^*$ are approximated by their thermal average values, which are determined by extremizing the
thermodynamic potential.

\begin{figure*}
\begin{center}
\includegraphics*[width=8.4cm]{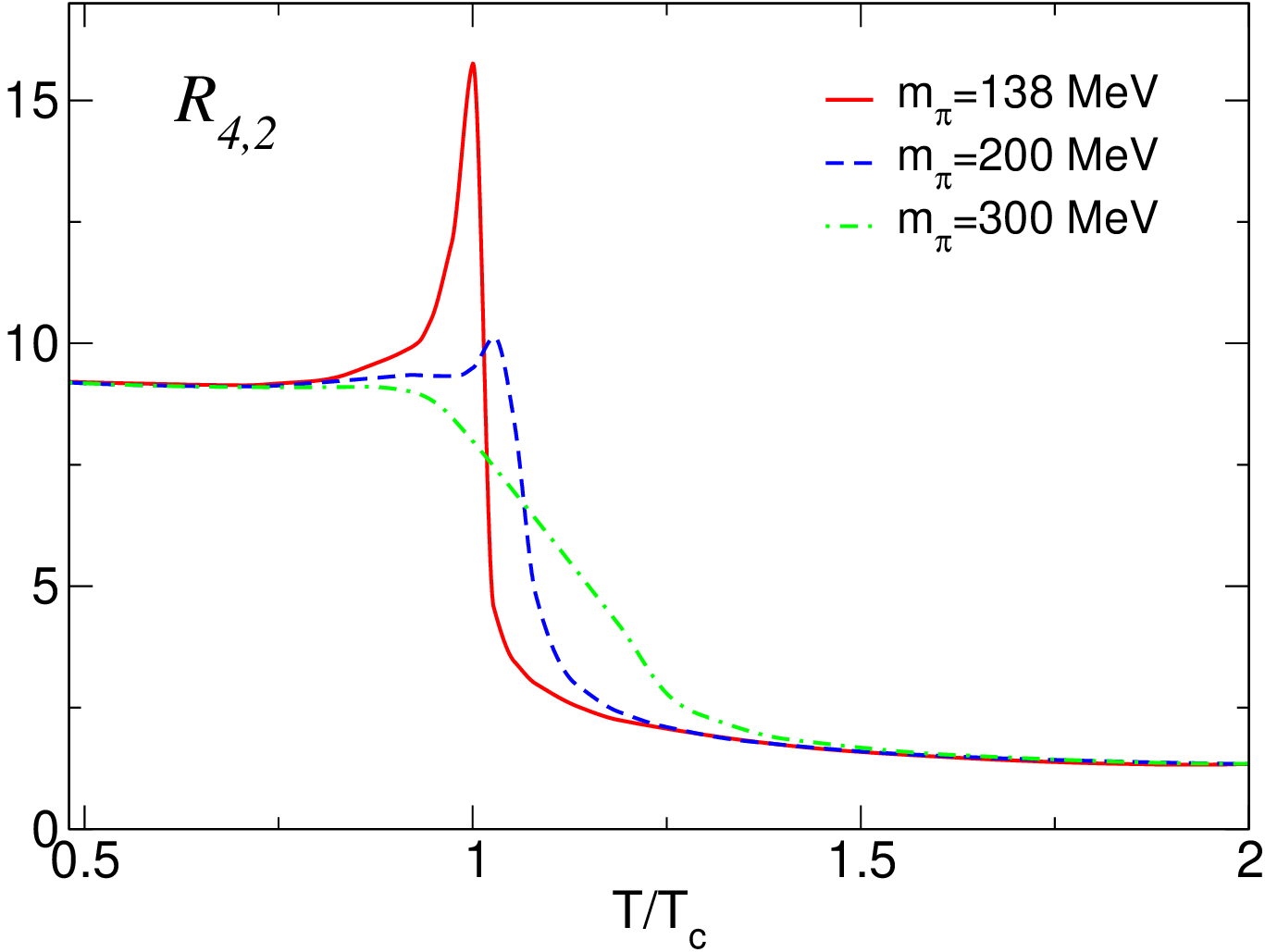}
\includegraphics*[width=8.15cm]{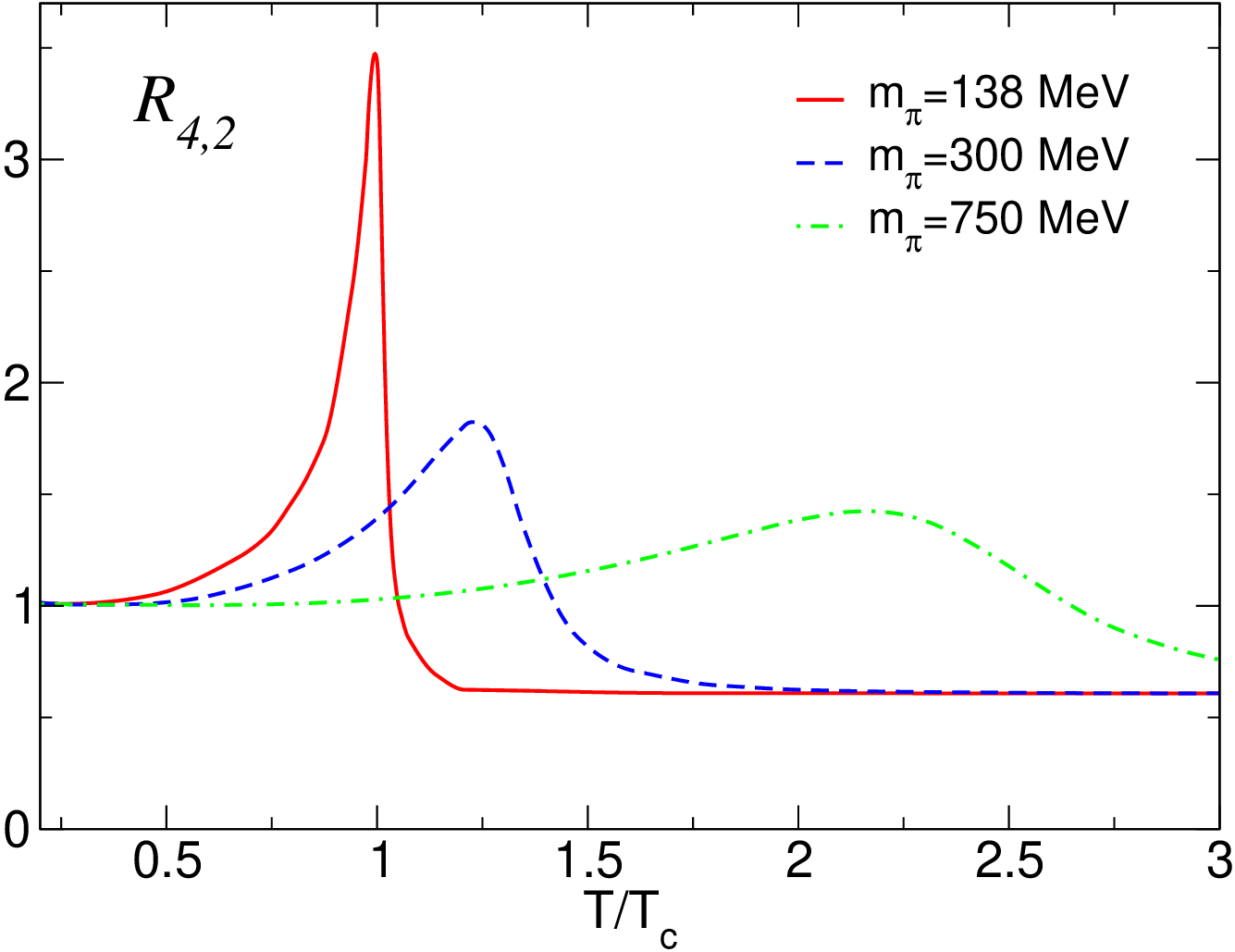}
\caption{The kurtosis $R_{4,2}$ for various values of the pion mass
in  the  PQM (left panel) and in the chiral quark--meson (QM) model (right  panel). The temperature is
normalized to the pseudocritical temperature $T_c$ for a physical value of the pion mass.  }\label{fig:ratio42_qm}
\end{center}
\end{figure*}

\section{Kurtosis and compressibility near the chiral transition}

In the following we focus on the quark sector of the thermodynamic potential. Following the procedure in LGT studies \cite{lattice:ejiri}, we use the Taylor expansion in $\mu/T$ to compute various observables. In particular the thermodynamic pressure is expressed as
\begin{equation}\label{eq:pressure}
 \frac{p\,(T,\mu)}{T^4}=\sum_{n=0}^{\infty}\frac{1}{n!}c_n(T)\left(\frac{\mu}{T} \right)^n
\end{equation}
with the expansion coefficients
\begin{equation}
\left. c_n(T)=\frac{\del^n(p\,(T,\mu)/T^4)}{\del(\mu/T)^n}\right\vert_{\mu=0}.
\end{equation}
The Taylor coefficients are generalized susceptibilities corresponding to moments of the net quark number
\cite{Allton:2003vx,Ejiri:2005wq,F1}. In particular, the first two non-vanishing derivatives,  $c_2$
and  $c_4$ are the second and fourth order cumulants which are related to fluctuations   of net quark 
number, $\delta N_q = N_q - \langle N_q\rangle$, as follows:

\begin{eqnarray}
c_2 &=& {\frac{\chi_q}{T^2}}=<(\delta N_q)^2> = <N_q^2>-<N_q>^2\quad
\nonumber \\
c_4 &=& <(\delta N_q)^4>-3<(\delta N_q)^2>^2    \quad , \label{fluctuations}
\end{eqnarray}
with $\chi_q$ being the quark number susceptibility. We focus on the  ratio
\begin{equation}\label{kar}
 R_{4,2}=\frac{c_4}{c_2},
\end{equation}
which we define as the  kurtosis of the quark number fluctuations (see footnote above). 
We consider the dimensionless ratio of the inverse compressibility and Gibb's free energy density $\mu n_q$,
\begin{equation}\label{kapa}
 R_{\kappa}=\frac{\kappa^{-1}}{\mu n_q}=\frac{n_q}{\mu \chi_q}\, ,
\end{equation}
where the inverse isothermal
compressibility  of the system is given by
\begin{equation}\label{compr}
\kappa^{-1} = - V\left(\frac{\partial p}{\partial V}\right)_T=\frac{n_q^2}{\chi_q}\, .
\end{equation}

Before considering  $R_{4,2}$ and $R_\kappa$  near the chiral phase transition,  we first focus on their behavior well above and below the transition region.
For high temperatures, i.e.  $T>>T_0$, the ratio $m_q/T<<1$ and the Polyakov loop $l\to 1$. Hence, the quark-antiquark contribution to the thermodynamic pressure (\ref{eq:qqbar_pot}) converges to that of an ideal gas of quarks, where
\begin{equation}\label{eq:free_pressure}
 \frac{P_{q\bar q}(T,\mu)}{T^4}={N_fN_c}
\left[\frac{1}{12\pi^2}\left(\frac{\mu}{T}\right)^4+\frac{1}{6}\left(\frac{\mu}{T}\right)^2+\frac{7\pi^2}{180}
\right].
\end{equation}
Thus, in the limit of high temperatures,   $R_{4,2}$ and  $R_\kappa$ are in both
models given by
\begin{equation}\label{a2}
\left.R_{4,2}\right\vert_{T>>T_0}=\frac{6}{\pi^2},\quad
\left.R_{\kappa}\right\vert_{T>>T_0}=\frac{(\mu/T)^2+\pi^2}{3(\mu/T)^2+\pi^2},
\end{equation}
where $R_{4,2}$ is evaluated at $\mu=0$, while for $R_\kappa$ we retain the $\mu$ dependence in anticipation of the application below.

One of the main features of a chiral Lagrangian coupled to the Polyakov loop is the "statistical
confinement" which implies, that at small $T$, the effective degrees of freedom are three quark states
\cite{Ratti:2006wg,Fukushima}. This is explicitly seen in Eq. (\ref{eq:qqbar_pot}), where  for
small $T<<T_0$ the Polyakov loop  $l\simeq  0$, implying suppression of the one- and two-quark states in the partition sum. Consequently, in the Boltzmann approximation\footnote{The
Boltzmann approximation is valid if $3m_q/T>1$. This condition is clearly satisfied for $T<T_0\simeq
200$ MeV since $m_q\simeq 300$ MeV.} the fermion contribution to the pressure
is
\begin{equation}\label{asp2}
 \frac{P_{q\bar q}(T,\mu)}{T^4}\simeq
 \frac{2N_f}{27\pi^2}\left(\frac{3m_q}{T}\right)^2K_2\left(\frac{3m_q}{T}\right)
\cosh{\frac{3\mu} {T}}.
\end{equation}
Thus, at low temperatures  the quark/antiquark contribution to the thermodynamic pressure is that of a
non-interacting gas composed of  particles and  antiparticles  with   mass $M=3m_q$ and baryon
number $B=1$ and $B=-1$, respectively. The effective degeneracy of these particles is
$2N_f/27$. The suppression by a factor $1/27$ is due to the rescaling of the momenta implied by the transformation $3\sqrt{p^2+m^2} \to \sqrt{k^2+(3m)^2}$.

An important feature of the pressure  (\ref{asp2}) is the factorization of the $M/T$  and $\mu/T$ dependence,
\begin{equation}\label{asp}
 \frac{P_{q\bar q}(T,\mu)}{T^4}=f(\frac {M}{T})\cosh{\frac{3\mu} {T}}\, .
\end{equation}
A similar factorization occurs in the hadron resonance gas model
HRG, although with a more complicated function replacing $f(M/T)$. In the HRG, the corresponding function involves a sum over all baryon states \cite{Ejiri:2005wq,T1,T2}. We note, however, that the Polyakov-loop models fail to reproduce the pressure of a resonance gas, even when the HRG is restricted to a gas of nucleons and $\Delta$ resonances (more generally the lowest three-quark multiplets) and their masses are approximated by $M_N\sim M_\Delta\sim 3m_q$.  The origin of this failure is twofold: on the one hand the suppression factor mentioned above and, on the other hand, an incorrect dependence on $N_f$. The latter is due to the fact that only states where all three quarks have the same flavor are counted in (\ref{asp2}). 

Since the HRG seems to provide a satisfactory description of the QCD equation of state at low temperatures~\cite{Ejiri:2005wq,T1,T2}, the effective model considered here can, strictly speaking, be valid only near the 
phase transition, where the dynamics is controlled by the chiral symmetry. However, the model may  also be
useful outside the critical region, for computing observables that are not sensitive to details of the mass spectrum. The kurtosis (\ref{kar}) and the
dimensionless inverse compressibility (\ref{kapa}) fulfil this criterion at least at low temperatures. Indeed, from Eqs. (\ref{kar}),  (\ref{kapa}) and
(\ref{asp}) it follows that the function $f(M/T)$, which depends on the mass spectrum, cancels
in the ratios, leading to 

\begin{equation}\label{a1}
\left.R_{4,2}\right\vert_{T<<T_0}=9,\quad
\left.R_{\kappa}\right\vert_{T<<T_0}=\frac{T}{3\mu}\tanh{\frac{3\mu}{T}}.
\end{equation}
For reference we also give the dimensionless inverse compressibility in the approximation, where $n_q$ and $\chi_q$ are computed to next-to-leading order in $\mu/T$
\begin{equation}\label{kappa-exp}
\left.R_{\kappa}\right\vert_{T<<T_0}=\frac{2+3(\mu/T)^2}{2+9(\mu/T)^2}\, .
\end{equation}

The kurtosis (\ref{kar}) is strongly dependent on the  net quark content of
the baryon number carrying effective degrees of freedom. For instance, at low temperatures, $R_{4,2}=(3B)^2$. Consequently, if the low-temperature phase is confining, $B=1$ and $R_{4,2}=9$, as found also in the hadron resonance
gas model~\cite{Allton:2003vx,Ejiri:2005wq}. 
On the other hand, in the pure Quark-Meson (QM) model, i.e. neglecting the effects of the Polyakov loop, the relevant fermionic degrees of freedom are single quarks, at all temperatures. Therefore, the corresponding value of the kurtosis in the low temperature limit is $R_{4,2}=1$. Clearly, in an effective chiral model with quark degrees of freedom, the Polyakov loop is essential for obtaining the correct low temperature behavior of the kurtosis~\footnote{We stress again that Polyakov loop models can reproduce the correct value of $R_{4,2}$, although they do not yield a realistic description of the pressure at low temperatures, because the dependence on details of the baryon spectrum cancels in the ratio $c_4/c_2$.}. 

We now turn to the temperature dependence of $R_{4,2}$ and $R_\kappa$. 
By inspection of Eqs. (\ref{a2}) and (\ref{a1}) we conclude that both the kurtosis and $R_\kappa$ are, for fixed $\mu/T$, temperature independent far above and below the transition. However, the high- and low-temperature values differ considerably, implying that the two ratios must vary near the phase transition. The major contribution to the change of the kurtosis $R_{4,2}$ between the low and high temperature limits is due to the change of the relevant baryon number carrying degrees of freedom, from three-quark states to single quarks.  Thus, indeed the kurtosis is, as noted in \cite{Ejiri:2005wq}, an
excellent probe of deconfinement. It is clear from this discussion that,
far above and below the transition, $R_{4,2}$ and $R_\kappa$  are independent of $m_\pi$. However, as illustrated in
Fig.~\ref{fig:ratio42_qm} this is not the case in the phase transition region.

In Fig.~\ref{fig:ratio42_qm} we also show, that for $T<T_{pc}$ 
the kurtosis in both the  PQM and QM models converges fairly rapidly to its  low-temperature limit, independently of the value of $m_\pi$. 
The temperature (in)dependence of $R_{4,2}$ in both models is consistent with LGT findings  in 2- and 2+1-flavor QCD \cite{Allton:2003vx,Ejiri:2005wq}. However, as discussed above, only the PQM model correctly reproduces
the low $T$ limit, $\left.R_{4,2}\right\vert_{T<<T_{pc}}=9$. 

\begin{figure*}
\begin{center}
\includegraphics*[width=8.6cm]{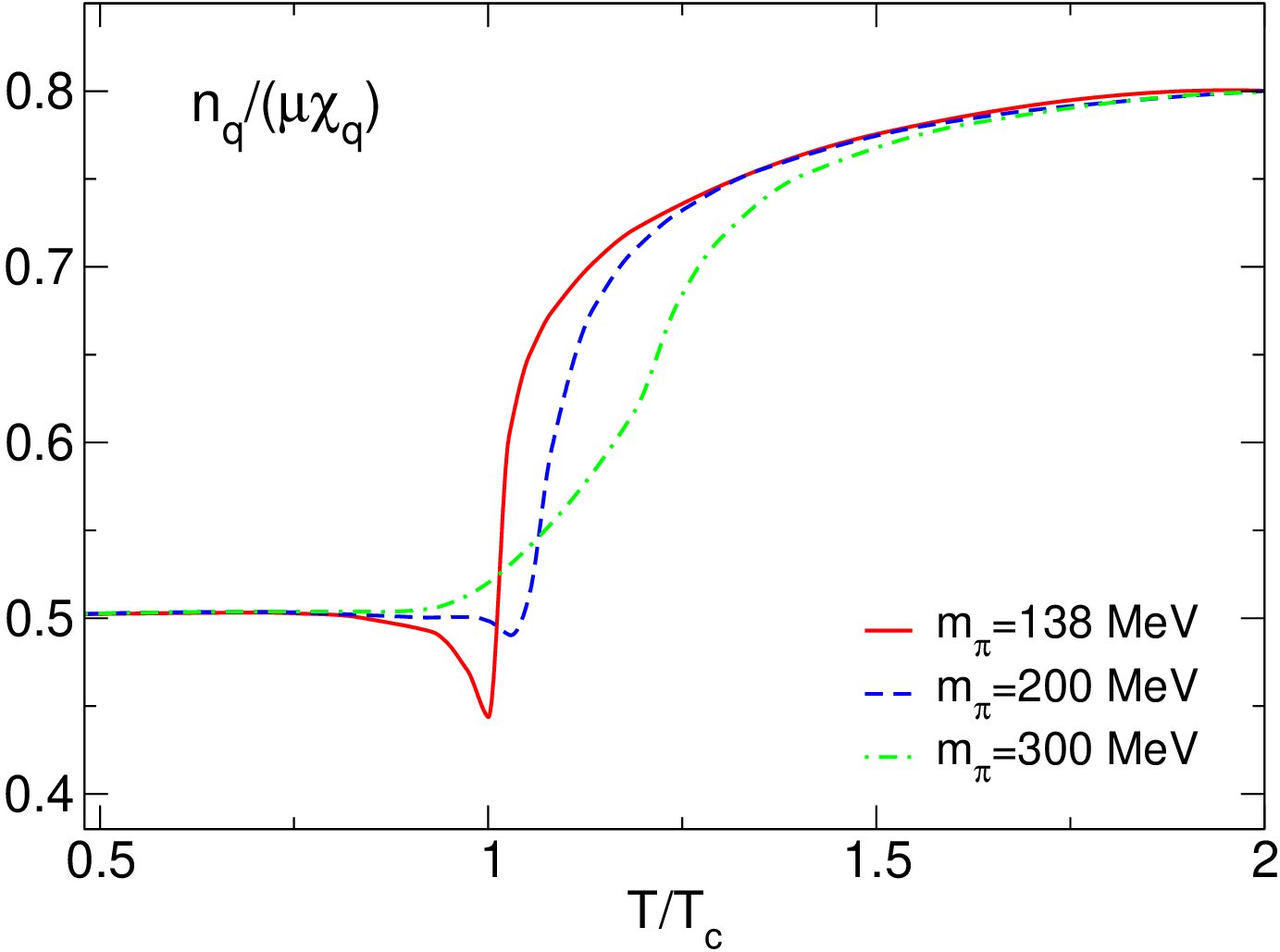}
\includegraphics*[width=8.6cm]{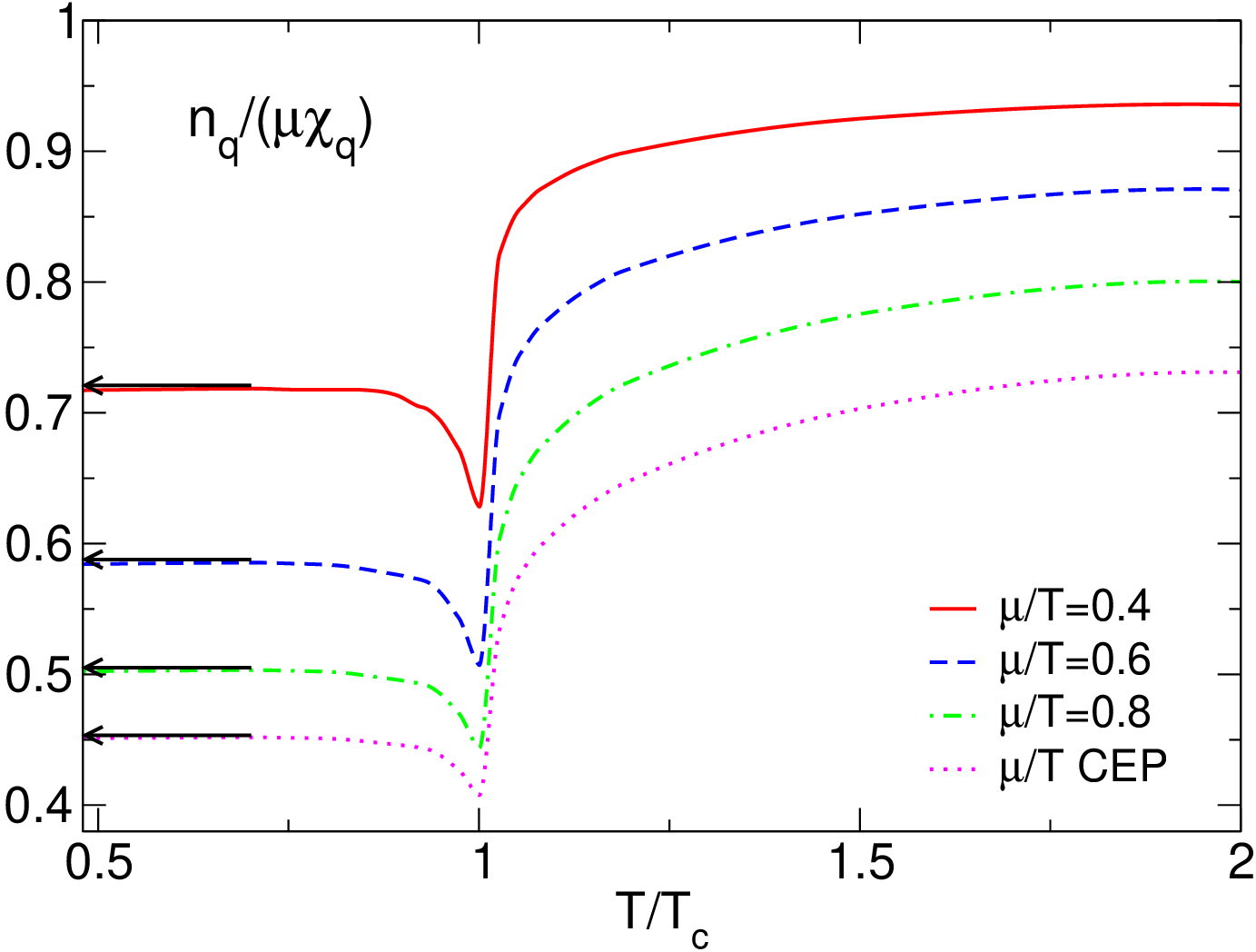}
\caption{The inverse dimensionless compressibility $R_\kappa$ computed in the PQM model at
fixed $\mu/T=0.8$ for various values of $m_\pi$ (left panel) and at fixed $m_\pi=138$ MeV for different
$\mu/T$ (right panel). The curve labelled CEP corresponds to the value of $\mu/T$ at the critical end point.  The arrows in the right panel  indicate the low-temperature values of $R_\kappa$
at a given $\mu/T$.
The temperature is normalized as in   the Fig. \protect\ref{fig:ratio42_qm}.
}
 \label{fig:ratio_n_chi_qm_pqm_pion}
\end{center}
\end{figure*}

At high temperatures, the kurtosis in the QM model reaches the asymptotic value already at $T\simeq 1.5T_{pc}$,
independently on the  value of the pion mass, in full agreement with LGT results
\cite{Allton:2003vx,Ejiri:2005wq}.  However, this is not the case for the PQM model, where even
at $T> 2 T_{pc}$, $R_{4,2}$ is still about $50\%$ above the ideal gas value. The
different high-temperature behavior of the  PQM and QM models is clearly a consequence of the coupling of the quarks to the Polyakov loop rather than an effect of the mean field dynamics, as conjectured in Ref.~\cite{Ghosh:2006qh}.
The ideal gas value of $R_{4,2}$ is in the PQM model reached at high $T$ if both $m_q/T<<1$ and
$l\simeq 1$. The first requirement is indeed satisfied already at $T>(1.2-1.5)T_c$, but the
second is not valid even at $T> 2 T_{pc}$. This is a direct consequence of the parametrization of the Polyakov
loop effective potential in the PQM Lagrangian (\ref{eq:pqm_lagrangian}). The parameters of the effective potential were fixed by fitting $SU(3)$ lattice results,
where $l$ approaches its asymptotic value very slowly. This discussion points to a clear limitation of
the applicability of the PQM model to the interpretation of LGT results  at high temperatures where e.g., the
interactions of quarks with space-like Wilson loops, not included in the PQM
model, could be relevant.

Near the chiral transition,  the kurtosis  is strongly dependent on the pion mass. For
a physical pion mass, $R_{4,2}$ exhibits a peak in both models.  With increasing $m_\pi$ the peak height gradually decreases and its position is shifted to larger temperatures, reflecting an increase in $T_{pc}$.  The dependence of the kurtosis on the pion mass is stronger in the QM than in the PQM model. A comparison with  LGT results \cite{Allton:2003vx, Ejiri:2005wq} show that the PQM
model yields a much better overall description of the LGT data. For large $m_\pi$ there is a smooth change of the
kurtosis between its limiting high- and low-temperature values, in agreement with 2-flavor QCD on the lattice \cite{Allton:2003vx, Ejiri:2005wq}. On the other hand,  for physical $m_\pi$, $R_{4,2}$ increases at $T_{pc}$ beyond the low-temperature value $R_{4,2}\simeq 9$. Both these
results, together with the low temperature limit, are consistent
with the LGT results in 2 and 2+1 flavor QCD, obtained for $m_\pi\simeq 770$ and $m_\pi\simeq
220$ MeV, respectively. The only problem of the PQM model in this context is the high-temperature behavior of $R_{4,2}$, discussed above, while the QM model yields an incorrect low-temperature limit and a too strong dependence on the pion mass near the phase transition. Consequently, we restrict our discussion below to the PQM model.

As a
consequence of  $O(4)$ scaling, one expects a strong variation of the kurtosis with $m_\pi$ near the chiral phase transition~\cite{Ejiri:2005wq}. At
$\mu=0$, the singular part of the net quark number fluctuations scales as $\chi_q\sim t^{1-\alpha}$
whereas that of the fourth order cumulant is of the form $c_4\simeq t^{-\alpha}$, where $t=|T-T_c|/T_c$ and $\alpha$ is the critical exponent of the
specific heat. In the $O(4)$ universality class, $\alpha$ is small and negative, $\alpha\simeq -0.26$. Consequently, the regular part dominates in the susceptibility $\chi_q$, whereas the singular part of $c_4$
corresponds to a cusp. In the presence of a term that explicitly breaks the $O(4)$ symmetry, the cusp is smoothened and can, for small pion masses, be seen in $R_{4,2}$ as a
peak at $T_{pc}$. For larger $m_\pi$, $c_4$ is dominated by the regular part, resulting in
monotonic change of $R_{4,2}$ in the transition region. These expectations, which are borne out in the PQM
model, are consistent with LGT results.

At finite chemical potential, the singularity is stronger; $c_4$ diverges at the $O(4)$ line whereas the quark number
susceptibility $\chi_q$ develops  a cusp. Furthermore, $\chi_q$ diverges at the
critical end point (CEP), in accordance with $Z(2)$ universality. Consequently, the inverse compressibility,  introduced in the  Eq. (\ref{kapa}), is an interesting  observable that
can be used to verify the existence of the CEP, to identify its position, provided it exists and to establish the universality class of the chiral transition.

In left panel of Fig.~\ref{fig:ratio_n_chi_qm_pqm_pion} we show the temperature dependence of the dimensionless inverse compressibility $R_\kappa$ (\ref{kapa})
computed in the PQM model for fixed $\mu/T=0.8$ and for various values of  $m_\pi$. We follow the procedure in the LGT calculations of~\cite{Allton:2003vx, Ejiri:2005wq}, and compute $n_q$ and $\chi_q$ keeping only the first two non-vanishing
coefficients, $c_2$ and $c_4$. The low $T$ values of $R_\kappa$ are consistent with the low-temperature limit (\ref{kappa-exp}), indicated by the arrows in the right panel.
For large temperatures, $T>>T_{pc}$, $R_\kappa$ is independent of $m_\pi$ and slowly converges  to the ideal
gas value (\ref{a2}). Similarly as in the case of  $R_{4,2}$,  $R_\kappa$ deviates substantially from
the  Stefan-Boltzmann result, due to the gentle approach of the Polyakov loop to unity.

Near the chiral transition, $R_\kappa$  shows a rather  strong dependence on the pion mass. For
large $m_\pi$, $R_\kappa$  is smoothly increasing between the low- and  high-temperature values. On the other hand, for a physical value of the pion mass ($m_{\pi}=138$ MeV), a dip develops in
the vicinity of $T_{pc}$. As shown in the right panel of Fig. \ref{fig:ratio_n_chi_qm_pqm_pion}, the value of $R_\kappa$ at the dip drops
with increasing $\mu/T$. This is consistent with recent lattice results~\cite{FN} obtained in
2+1 flavor QCD for $m_\pi=220$ MeV. The agreement of the two-flavor PQM model results with those of 2+1 flavor 
LGT suggests that the lattice results reflect the critical behavior of $O(4)$ chiral dynamics.

For the parameters used in our calculations, the CEP appears in the PQM model at $\mu/T\simeq 1$. Consequently, at this value of $\mu/T$, $R_\kappa$ should vanish at the
critical temperature. It is clear from the right panel of Fig.~\ref{fig:ratio_n_chi_qm_pqm_pion} that $R_\kappa$
clearly reflects the $O(4)$ critical dynamics. However, the approximate $R_\kappa$, obtained by a Taylor expansion of the numerator and the denominator,   is not adequate  to
identify the position of the CEP by a zero of $R_\kappa$. Although the value of  $R_\kappa$ at $T_{pc}$
drops with increasing $\mu/T$, as the CEP is approached, it remains different from zero.
This is clearly a consequence of the Taylor expansion, since the zero of $R_\kappa$ is due to the singularity  
of the quark number susceptibility, which is not reproduced by a polynomial of finite order.


\section{Conclusions}

Based on the quark-meson chiral Lagrangian coupled to the Polyakov loop we have explored the
properties of the kurtosis $R_{4,2}$ of the net quark number fluctuations and the dimensionless inverse
compressibility $R_\kappa$.  We have focused on the behavior of these observables
near chiral transition as functions of the pion mass.

We have  shown, that it is necessary to include the  coupling of the quarks to the Polyakov loop, in order to reproduce the low temperature behavior of the kurtosis expected in QCD. 
We have also demonstrated that near the pseudocritical
temperature, both these observables are very sensitive to the pion mass. For a physical value of $m_\pi$, the kurtosis
develops a peak and the inverse compressibility a dip at the pseudocritical temperature. These
structures disappear for large  $m_\pi$. Our results supports the notion that, for physical quark masses, the strange quark decouples from the critical fluctuations and the scaling properties are governed by $O(4)$, the expected universality class of two
flavor QCD. We also found a suppression of $R_\kappa$, as the critical end point CEP is approached. However, $R_\kappa$ remains finite at the CEP,  if  the Taylor
expansion in $\mu/T$ of the thermodynamic pressure is implemented.

\vspace*{-0.1cm}

\section*{ Acknowledgments}
K.R. acknowledges stimulating discussions with F. Karsch and the Organizers of the  INT "QCD Critical
Point" meeting where this paper was initiated. K.R. also acknowledges partial support from the
Polish Ministry of Science and the Deutsche Forschungsgemeinschaft (DFG) under the Mercator
Programme. B.S. gratefully acknowledges financial support from the Helmholtz Research School on
Quark Matter Studies.


\end{document}